\documentclass[12pt]{amsart}
\usepackage{amsmath, amssymb, amsthm}
\usepackage{mathabx} 
\usepackage{graphicx}
\usepackage[left=1.25in,top=1.4in,right=1.25in,bottom=1.4in,head=0.5in,foot=0.5in]{geometry}
\usepackage[breaklinks]{hyperref}
\usepackage{natbib}
\setcitestyle{aysep={}}
\usepackage[nodisplayskipstretch]{setspace}
\newtheorem*{genspecthm}{Naimark Spectral Theorem}
\newtheorem*{genstonethm}{Generalised Stone Theorem}
\theoremstyle{definition}
\newtheorem*{fact}{Fact}
\newtheorem{prop}{Proposition}

\makeatletter
\let\uppercasenonmath\@gobble

\makeatother

\newcommand{\CC}{\mathbb{C}}
\newcommand{\RR}{\mathbb{R}}
\newcommand{\HH}{\mathcal{H}}
\newcommand{\Inn}[1]{\langle #1 \rangle}
\newcommand{\Tr}[1]{\mathrm{Tr}(#1)}
\newcommand{\mymatrix}[2]{\begin{pmatrix} #1 \\ #2 \end{pmatrix}}
\newcommand{\smmatrix}[2]{\bigl(\begin{smallmatrix}#1 \\ #2
\end{smallmatrix}\bigr)}

\thanks{\emph{Acknowledgements.} Thanks to Michel Janssen for helpful comments.}

\title{Observables, Disassembled}
\author[Bryan W. Roberts]{Bryan W. Roberts}
\begin{document}
\maketitle
\begin{abstract}
This paper argues that non-self-adjoint operators can be observables. There are only four ways for this to occur: non-self-adjoint observables can either be normal operators, or be symmetric, or have a real spectrum, or have none of these three properties. I explore each of these four classes of observables, arguing that the class of normal operators provides an equivalent formulation of quantum theory, whereas the other classes considerably extend it.
\end{abstract}

\section{Introduction}

\begin{figure}[h!bt]\begin{center}
    \includegraphics[width=0.5\textwidth]{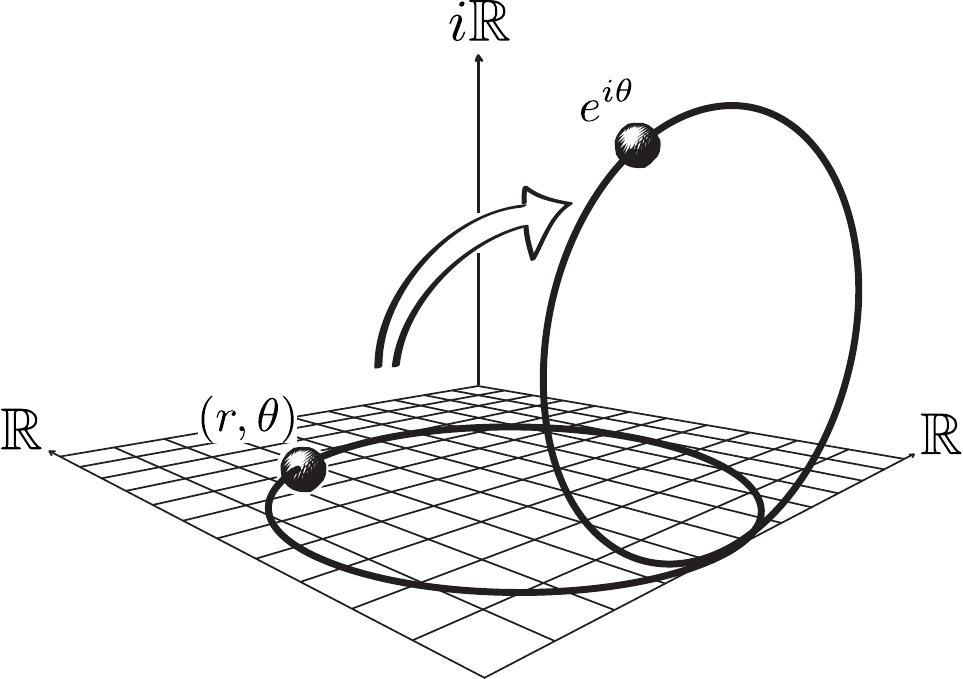}
    \caption{Real and complex descriptions of particle position related by a rotation in $\CC^2$.}\label{fig:ring}
  \end{center}\end{figure}

There is nothing wrong with using complex numbers to represent the physical world. Consider a bead that is constrained to move on a ring. If the members of some primitive species were to describe the position of that bead in space, they could use four arbitrary squiggles like $\Saturn$, $\Uranus$, $\Neptune$, $\Pluto$ to represent East, North, West and South. Or, once they constructed the real number field, they could use coordinates $(r,\theta)$ on a plane, where $r$ is the distance from some center point and $\theta\in[0,2\pi)$ is the angle from the right horizontal. Or they could adjoin an imaginary number $i$ and use the numbers on the complex unit circle $e^{i\theta}\in\mathbb{C}$ with $\theta\in[0,2\pi)$. The latter is no less reasonable than the other two as a representation of the bead's position. Of course, there was once considerable skepticism about the status of complex numbers\footnote{Cardano derived complex solutions to the equation $x^2 - 10x + 30 = 0$ in his 1545 \emph{Ars Magna}, but concluded, ``So progresses arithmetic subtlety the end of which, as is said, is as refined as it is useless'' \cite[\S 37]{cardano-arsmagna}. Over 200 years later Euler took a similar view: ``they are usually called \emph{imaginary quantities}, because they exist merely in the imagination'', although he argued that ``nothing prevents us from making use of these imaginary numbers, and employing them in calculation'' \citep[p.43]{euler-algebra}.}. But such misgivings need not trouble us today: the complex numbers can be constructed axiomatically in just the same sense as the real numbers, and provide an equally adequate labeling scheme for the position of the bead. Indeed, if we view the real-number loop and the complex-number loop as both embedded in the complex plane $\CC^2$, then the two descriptions are related by a rotation, as shown in Figure \ref{fig:ring}.

I would like to argue that this egalitarian perspective should be applied to quantum theory as well. Orthodox quantum theory has long presumed that the outomes of experiment must be represented by the eigenvalues (and more generally the spectrum) of a self-adjoint operator, which are always real numbers. An operator interpreted in this way is then dubbed an `observable'. But it does not have to be this way: there are many physically and philosophically interesting ways to have a non-self-adjoint observable.

To see this, I suggest we break the property of self-adjointness into three `component' properties: being normal, being symmetric, and having a real spectrum, which I define precisely below. My thesis is that observables can be represented by non-self-adjoint operators that have any one of these properties while giving up the other two, or that give up all three. It turns out that this exhausts the possible ways to have a non-self-adjoint observable.

The plan of the paper is as follows. The second section will introduce the dogma of self-adjoint operators, and then propose a way to classify the possible non-self-adjoint observables. The third section considers non-self-adjoint operators that are normal. Here I formulate a sense in which normal operators provide an equivalent framework for doing quantum mechanics. The fourth section explores the physics of non-normal operators: first operators that are symmetric but not real, which allow for the introduction of `time observables'; next operators that are real but not symmetric, which give rise to $PT$-symmetric observables; and finally operators that are neither normal, real nor symmetric. The fifth section is the conclusion.

\section{Self-adjointness disassembled}

\subsection{The rise of self-adjointness}

Heisenberg arrived in G\"ottingen in June of 1925 with a draft of his celebrated paper on non-commutative mechanics. But it was Max Born who famously recognised upon seeing this draft that the theory could be represented in terms of matrices. Soon, \citet{born1925quantenmechanik} had formulated the observables of quantum mechanics as self-adjoint or `Hermitian' operators\footnote{Charmingly, their collaboration apparently began by chance, on a train to Hanover soon after Born met Heisenberg in 1925. Born recalls confiding to a colleague on the train that he had formulated Heisenberg's equations of motion using matrix theory, but was stuck trying to derive the energy from this. Jordan, who was sitting opposite and overheard the conversation, said, ``Professor, I know about matrices, can I help you?' Born suggested they give it a try, and a famous collaboration ensued \citep[from an interview with Born by][]{Born-EwaldInterview-i}.}. In a letter to Jordan in September of that year, Heisenberg wrote,
\begin{quote}
  ``Now the learned G\"ottingen mathematicians talk so much about Hermitian matrices, but I do not even know what a matrix is\footnote{Quoted from \citet[p.207]{jammer-conceptualQM}. The impressive list of `learned mathematicians' at G\"ottingen when Heisenberg arrived in 1925 includes Paul Bernays, Max Born, Richard Courant, David Hilbert, Pascual Jordan, Emmy Noether, Lothar Nordheim, B.L. Van der Waerden, and Hermann Weyl.}.''
\end{quote}
As Heisenberg's letter reveals, matrices were far from common tools among physicists at the time, let alone Hermitian ones, despite the latter having been introduced by \citet{hermite1855remarque} seventy years earlier.

Physically significant non-Hermitian matrices appeared the following May, when \citet{london1926jacobian} derived the non-Hermitian raising and lowering operators for the harmonic oscillator. By December of 1926, \citet{jordan1927neueI} was actually toying with the idea of treating non-Hermitian operators as observables. Remarkably, Jordan's formalism allowed one to assign complex expectation values to such non-Hermitian operators, as \citet[\S 2.4]{DuncanJanssen2013never} have shown.  But in April of 1927, Hilbert, von Neumann and Nordheim had identified self-adjoint operators as appropriate for ensuring that the values of energy are always positive numbers\footnote{\citep{hilbert1928grundlagen}. As Janssen and Duncan point out, this article was submitted in April 1927, but ``for whatever reason'' not published until 1928 \citep[\S 3, p.221]{DuncanJanssen2013never}.}. By the time \citet{jordan1927neueII} submitted a follow-up paper in June, he had given up on the idea of non-Hermitian observables idea in favour of the new dogma\footnote{See \citet{DuncanJanssen2009jordan,DuncanJanssen2013never} for a fascinating exposition of this episode in the development of quantum mechanics.}.

Like many aspects of quantum theory as we know it, self-adjointness was consolidated at the September 1927 Solvay conference, where Born and Heisenberg's report argued that,
\begin{quote}
``the analogy with classical [Fourier] theory leads further to allowing as representatives of real quantities only matrices that are Hermitian'' \citep[p.327]{BornHeisenberg1927a}
\end{quote}
Their idea is a familiar one: it is often convenient to use a complex unit $e^{i\theta}=A\cos\theta + iA\sin\theta$ to represent a harmonic phenomenon like a classical wave, on the understanding that the amplitude and position of a physical wavecrest is described by just the real part, $\mathrm{Re}( e^{i\theta})=A\cos\theta$.

The dogma soon became encoded in the influential textbooks of the field, including Dirac's famous \emph{Principles of Quantum Mechanics}. In the 1930 first edition, Dirac actually used the term `observables' to refer to all linear operators, arguing that the algebra of observables for quantum mechanics involves ``an extension of the meaning of observable to include the analogues of complex functions of classical dynamical variables'' --- but was quick to add that ``[a]n observable is thus not necessarily capable of direct measurement by a single observation, but is a theoretical generalization of such a quantity'' \citep[p.27-28]{dirac1930principles}. Dirac revised this language in the second edition five years later, writing, ``it is preferable to restrict the word `observable' to refer to real functions of dynamical variables and to introduce a corresponding restriction on the linear operators that represent observables'' \citep[p.29]{dirac1935principles}. The `corresponding restriction' was that observables be self-adjoint.

Dirac's dictum has continued to be a pervasive dogma in modern textbooks, with an emphasis on the fact that they have a real spectrum. Griffiths justified the self-adjointness property by arguing, ``the expectation value of an observable quantity has got to be a real number (after all, it corresponds to actual measurements in the laboratory, using rulers and clocks and meters)'' \citep[\S 3.3]{griffiths1995qm}. Similarly, Sakurai wrote, ``[w]e expect on physical grounds that an observable has real eigenvalues... That is why we talk about Hermitian observables in quantum mechanics'' \citep[\S 1.3]{sakurai1994}. And Weinberg remarked in his \emph{Lectures on Quantum Mechanics} that, ``[w]e can now see why it is important for all operators representing obseravble quantities to be Hermitian. ... Hermitian operators have real expectation values'' \citep[p.24]{weinberg2013a}.

As one might expect, the philosophy of quantum mechanics has largely followed the textbooks. For example, David Albert's book on the philosophy of quantum mechanics first sets out what he calls `principle (B)', that measurable properties are to be represented by linear operators, and then states, ``it's clear from principle (B) (since, of course, the values of physically measurable quantities are always real numbers) that the operators associated with measurable properties must necessarily be Hermitian operators'' \citep[p.40]{albert1992}. Similar remarks are found in many other places in physics an philosophy.

I will argue that the dogma requiring self-adjoint quantum observables should be abandoned. To be precise about what I'm advocating, let me begin by setting out a few mathematical definitions and prerequisites that these discussions are not always sensitive to.

\subsection{Mathematical prerequisites}

This discussion will deal entirely with Hilbert spaces over the complex field that admit a countable (though possibly infinite) basis. Some of the Hilbert space operators we discuss will be unbounded, which implies that their domains\footnote{The \emph{domain} $D_A$ of an operator $A$ on a Hilbert space $\HH$ is the set of vectors $\psi$ such that $A\psi\in\HH$.} are not equal to the entire Hilbert space. When that is the case, I will still presume that they are at least densely-defined and closed\footnote{An operator $A$ is \emph{densely-defined} iff its domain $D_A$ is dense; this assures that the operator is minimally well-defined on `most' vector states. It is \emph{closed} iff for any sequence $\{\phi_n\}\subseteq D_A$ such that $\phi_n \rightarrow \phi$ and $A\psi_n\rightarrow \psi$, it follows that $\phi\in D_A$ and $\psi=A\phi$. This assures that the spectrum is non-trivial; if a densely defined operator is \emph{not} closed then its spectrum is $\mathrm{Sp}(A)=\mathbb{C}$.}. I will write $A^*$ to denote the adjoint\footnote{The \emph{adjoint} of $A$ is defined by $A^*\psi:=\psi^*$, where $\Inn{\psi^*,\phi} = \Inn{\psi,A\phi}$ for all $\phi$ in the domain of $A$. The domain of $A^*$ consists of those vectors $\psi$ for which such an element $\psi^*$ exists.} (or `conjugate transpose') of $A$. An operator $A$ is called \emph{normal} if it commutes with its adjoint, $AA^* = A^*A$. It is \emph{symmetric} if it has the property that $A\psi=A^*\psi$ for all $\psi$ in the common domain of $A$ and $A^*$. It is \emph{self-adjoint} if it is both symmetric and has the property that the $A$ and $A^*$ have the same domain\footnote{Every symmetric operator satisfies $D_A\subseteq D_{A^*}$. So, the additional condition that $D_A=D_{A^*}$ is equivalent to the statement that $D_{A^*}\subseteq D_{A}$. See \citet[\S 4]{BlankExnerHavlicek}.}. The term `Hermitian' is sometimes used for one or both of these last two properties; this is unambiguous if $A$ is bounded, in which case an operator is symmetric if and only if it is self-adjoint. But since this equivalence fails for unbounded operators, I will try to reduce confusion by avoiding the term `Hermitian'.

The \emph{spectrum} of a linear operator $A$ is the set of numbers $\lambda\in\mathbb{C}$ such that the operator $(A-\lambda I)$ does not admit an inverse. The \emph{eigenvalues} of $A$ are the subset of the spectrum consisting of elements $\lambda$ that satisfy $A\psi = \lambda\psi$ for some $\psi$. We say that an operator has a \emph{discrete} or \emph{pure point} spectrum when its spectrum consists entirely of eigenvalues. All operators on a finite-dimensional Hilbert space have a discrete spectrum, but the spectrum may contain elements that are not eigenvalues in the infinite-dimensional case. Finally, an important fact for our discussion is that in general, if $A$ is self-adjoint, then its spectrum (and thus its set of eigenvalues) is entirely composed of real numbers.

How do normal operators, symmetric operators, and operators with a real spectrum underpin the property of self-adjointness? The answer is given by following.

\begin{figure}[tbh]\begin{center}
    \includegraphics[width=0.4\textwidth]{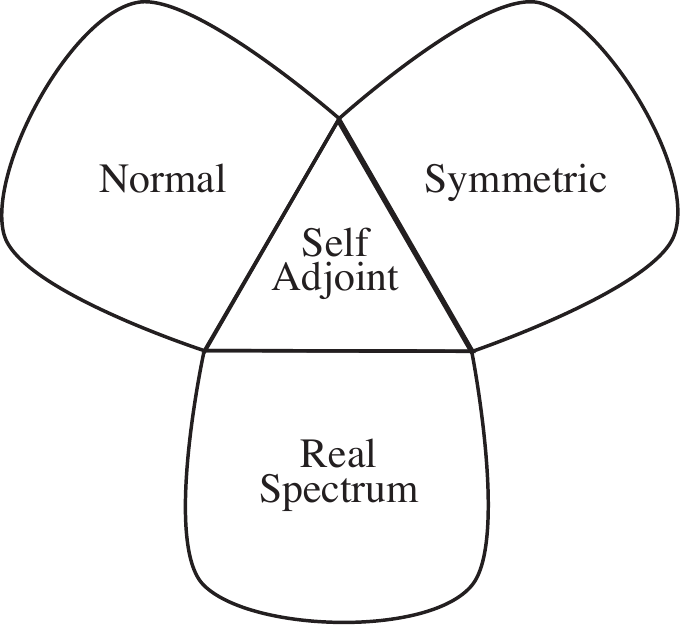}
    \caption{Venn diagram of normal, symmetric, and real-spectrum operators, any two of which imply self-adjointness. For bounded operators, being symmetric is equivalent to being self-adjoint, and so the right `petal' vanishes.}\label{fig:sa}
\end{center}\end{figure}

\begin{fact}
A closed, densely-defined linear operator $A$ is self-adjoint if it satisfies any two of the following properties.
  \begin{enumerate}
      \item \emph{Normal.} $AA^* = A^*A$.
      \item \emph{Symmetric.} $A\psi = A^*\psi$ for all $\psi\in D_A$.
      \item \emph{Real spectrum.} $\mathrm{Sp}(A) \subseteq \mathbb{R}$.
  \end{enumerate}
Conversely, every self-adjoint operators satisfies all three of the properties above.
\end{fact}
This conveniently summarises several standard results\footnote{A normal operator is symmetric if and only if it is self-adjoint \citep[Thm. 4.3.1]{BlankExnerHavlicek}; a normal operator has a real spectrum if and only if it is self-adjoint \citep[Thm. 12.26]{rudin-functional}; and a symmetric operator has a real spectrum if and only if it is self-adjoint \cite[p.136, Thm. X.1(3)]{ReedSimon1975}.}. Note that no single one of the properties above is sufficient to guarantee that $A$ is self-adjoint: A normal operator can fail to be symmetric; a unitary operator is an example. A symmetric operator that is unbounded can fail to be normal; the so-called `maximal symmetric' operators (operators with no self-adjoint extension) are an example. And an operator with a real spectrum can fail to be symmetric. We will discuss more concrete examples of such operators over the course of this paper. But to keep the facts in one's head, it is helpful to refer to Figure \ref{fig:sa}.

Since many find the last property particularly surprising, that a non-self-adjoint operator may have a real spectrum, let me give a concrete example. A particularly simple one is the $2\times 2$ matrix,
\[
  A = \mymatrix{1 & 1}{0 & 2}.
\]
It is obviously not symmetric (and thus not self-adjoint), since the conjugate-transpose is given by $A^* = \smmatrix{1 & 0}{1 & 2} \neq A$. But one can easily check that it has exactly two eigenvalues, both of which are real: $\lambda = 1$ with eigenvector $\smmatrix{1}{0}$, and $\lambda=2$ with eigenvector $\smmatrix{1}{1}$. As expected, this operator fails to be normal, as one can verify by checking $AA^*\neq A^*A$. It also has the property that its eigenvectors span the Hilbert space, but are not orthogonal.

\subsection{A classification of non-self-adjoint observables}

This mathematical discussion suggests a classification scheme for non-self-adjoint observables. A consequence of the fact above is that all of the non-self-adjoint operators (that are closed and densely defined) fall into exactly one of the following four categories.
\begin{enumerate}
  \item \emph{Normal operators} that are non-symmetric and have non-real spectra;
  \item \emph{Symmetric operators} that are not normal and have non-real spectra;
  \item \emph{Real-spectrum operators} that are not normal and not symmetric;
  \item \emph{None of the above operators} that fail to have all three of these properties.
\end{enumerate}
That is, one can allow non-self-adjoint observables to include operators from exactly one of the three `petals' in the flower of Figure \ref{fig:sa}, or none of them. Note that if one restricts attention to bounded operators, then the symmetric petal vanishes, since for bounded operators being symmetric is equivalent to being self-adjoint.

I will discuss each of these four classes of non-self-adjoint observables in turn. They introduce varying degrees of conceptual difficulties, but I will identify circumstances in which each of them are reasonable. In the next section I will argue that the first class, the non-self-adjoint operators that are normal, actually provides a formalism that is equivalent to orthodox quantum theory with self-adjoint observables. The following section will then discuss how non-normal observables truly extend the theory.

\section{Normal operators as observables}

A simple example of a normal operator with a pure imaginary spectrum is $iQ$, where $Q$ is the position operator in the Schr\"odinger representation. It obviously commutes with its adjoint $(iQ)^* = -iQ$, and so it is normal. Its spectrum is a line in the complex plane (namely, the pure imaginary axis) and so it can be used to represent the position of a bead in one dimension of space. It even satisfies a natural commutation relation: if we represent momentum by $iP$, then $[iQ,iP]\psi = -[Q,P]\psi = -i\psi$ (working in units of $\hbar=1$). Another example is the unitary operator $e^{iQ}$, which also commutes with its adjoint, and has a spectrum equal to the complex unit circle. It can be used to represent the position of a bead on a loop depected in Figure \ref{fig:ring}. And it too can be given a natural commutation relation\footnote{One could simply take it to be given by the canonical commutation relations in Weyl form, $e^{iaP}e^{ibQ} = e^{iab}e^{ibQ}e^{iaP}$. \citet{levyleblond1976a} suggests an alternative expressed in terms of angular momentum.}.

My main argument in this section is that normal operators can be adopted as observables without losing any of the ordinary structure of quantum mechanics. Others have suggested this as well\footnote{See especially \citet{levyleblond1976a}, \citet[p.539]{penrose-roadtoreality}, and \citet[\S 2.4]{DuncanJanssen2013never}; this latter paper shows that normal operators can be used to formalise Jordan's early theory of non-self-adjoint observables.}, but I will try to give a systematic argument. I begin by identifying how one can still apply the statistical rules of quantum theory for normal operators, then point out a sense in which the normal operators are `projectively equivalent' to the self-adjoint operators, and finally discuss how symmetries and unitary evolution appear when normal operators are observables. 

\subsection{Interpreting observables}

\citet[\S 6]{reichenbach-qm} liked to reserve the term `observable' for things that can be directly verified using human sense organs, such as the positions of the spectral lines produced by a light source. He preferred the term `phenomena' for occurrences that might be only indirectly observed, like the emission of a photon from a Hydrogen atom, and `interphenomena' for everything in between. I prefer a simpler principle for thinking about observables in quantum theory: \emph{observables associate experimental states with symbols}. This has the advantage of being applicable at any of Reichenbach's levels, and makes the relationship between experiment and language more explicit. For example, when we say that an observable is represented by a discrete self-adjoint operator $A$, we are indicating an association between a set of experimental states, represented by the eigenstates $\varphi$ of $A$, with a set of symbols, the real-number eigenvalues of $A$. We typically use the latter to express quantitative facts about a magnitude registered by a physical device.

\begin{figure}[tbh]\begin{center}
\includegraphics[width=0.7\textwidth]{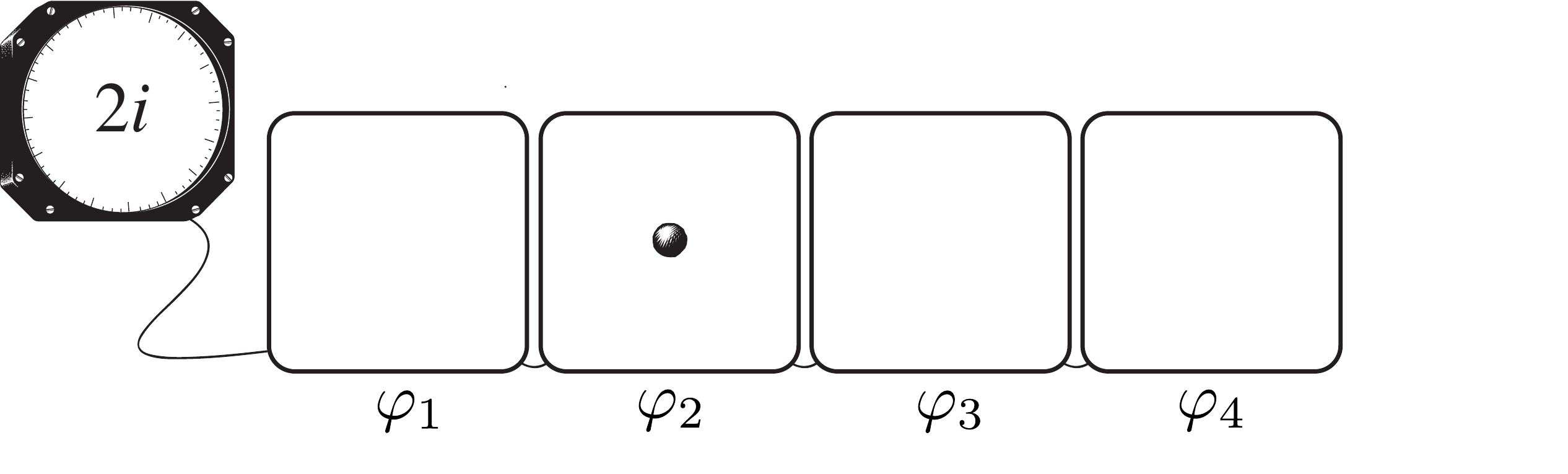}
    \caption{Eigenstates representing four distinct position states. The eigenvalue registered when the state $\varphi_n$ occurs is the complex number $(n)i$.}\label{fig:boxes}
  \end{center}\end{figure}

Notably, nothing about this practice requires the symbols to be real numbers; quantitative information can be conveyed by complex numbers as well, and by many other structures. Toy experiments describing observables with complex-number eigenvalues are easy to devise. For example, an experiment might be set up to detect a particle in exactly one of four definite locations, $\varphi_1$, $\varphi_2$, $\varphi_3$ or $\varphi_4$. A detector could display the symbols $i$, $2i$, $3i$, and $4i$ in each of these situations, respectively, as illustrated in Figure  \ref{fig:boxes}. The result is an association of definite states with symbols, namely complex numbers, which defines an observable in the general sense that I propose. 

The statistics for such an experiment can be defined just as they are in orthodox quantum mechanics: let $A$ be an operator on a Hilbert space of finite dimension, with a complex eigenvalue $\lambda$ corresponding to the eigenstate $\varphi$. Then the transition probability from an arbitrary state $\psi$ to $\varphi$ is still given by the usual Born rule, $|\Inn{\varphi,\psi}|^2$. If the eigenvectors of $A$ form a complete basis, then its expectation value when the state $\psi$ is prepared will still be given by $ \Inn{\psi,A\psi} = \sum_{i=1}^n \lambda_i|\Inn{\varphi_i,\psi}|^2$. For normal operators, such an expectation value may be a complex number, but this still makes good conceptual sense: a complex expectation value is just a weighted average of the complex numbers representing these states.

To confirm that the practice I am proposing has the same statistical interpretation as orthodox quantum theory, we also rely on the \emph{spectral theorem}. This is expressed in terms of a projection valued measure (or `spectral' measure) on Borel sets\footnote{A \emph{projection valued measure} on Borel subsets of a topological field $\mathbb{F}$ is a map $\Delta\mapsto E(\Delta)$, which associates each Borel subset $\Delta$ of $\mathbb{F}$ with a projection operator $E(\Delta)$, where $E(\mathbb{F})=I$ and $E(\bigcup_i \Delta_i) = \sum_i E(\Delta_i)$ for any countable disjoint collection $\{\Delta_1, \Delta_2, \dots\}$ that weakly converges. It follows from this that $E(\varnothing) = 0$, and $E(\Delta_1)E(\Delta_2)=0$ for disjoint $\Delta_1$, $\Delta_2$.}. In its statement for (possibly unbounded) self-adjoint operators, it says that a self-adjoint operator $A$ admits a unique projection valued measure $\Delta\mapsto E_\Delta$ on Borel sets of the reals such that $A = \int_{\mathbb{R}}\lambda dE_\lambda$ \citep[Theorem 5.3.1]{BlankExnerHavlicek}. In finite dimensions, the integral gets expressed as the sum,
\[
  A=\sum_i^n \lambda_i E_i,
\]
where each $\lambda_i$ is a real-number eigenvalue of $A$, and the projections $E_i$ satisfy $\sum_i^n E_i=1$, and also $E_iE_j=0$ when $i\neq j$. One of the conceptually important consequences of this theorem for quantum theory is that it allows us to view each state as defining a probability distribution on definite experimental outcomes associated with $A$. For example, in the finite-dimensional case, the spectral theorem implies there is a set of orthogonal, unit-norm eigenvectors $\varphi_1, \varphi_2, \dots, \varphi_n$ of $A$ that form a basis for the Hilbert space. That fact is what allowed Born to view a vector $\psi$ as defining a probability distribution $p_\psi(\varphi_i) := |\Inn{\varphi_i,\psi}|^2$, since it implies $\sum_i^n p_\psi(\varphi_i)=1$. Messiah thus writes in his classic textbook that, 
\begin{quote}
   ``[a]ll... operators do not possess a complete, orthonormal set of eigenfunctions. However, the Hermitian operators capable of representing physical quantities possess such a set. For this reason we give the name `observable' to such operators'' \citep[\S V.9]{messiah1999}.
\end{quote}

But in fact, by Messiah's reasoning, we should give the name `observable' to normal operators, too! All normal operators possess a `complete, orthonormal set of eigenfunctions' of the kind Messiah demands. This is guaranteed by a generalisation of the spectral theorem. It says that if $N$ is a normal operator, then there is a unique projection valued measure $\Delta\mapsto E_\Delta$ on Borel sets of $\mathbb{C}$ such that $N = \int_{\mathbb{C}}\zeta dE_\zeta$ \citep[Theorem X.4.11]{conway-functional}. In finite dimensions, this gets expressed as,
\[
  N = \sum_i^n \zeta_i E_i,
\]
where each $\zeta_i$ is a complex-number eigenvalue of $N$, and where the properties of the projections $E_i$ carry over exactly as in the self-adjoint case. This means that, just as with self-adjoint operators, every state defines a probability distribution on the experimental outcomes associated with a normal operator $N$. And just as with self-adjoint operators, a normal operator $N$ in finite dimensions has a set of orthonormal eigenvectors that form a basis for the Hilbert space, with $p_\psi(\varphi_i) := |\Inn{\varphi_i,\psi}|^2$ defining a probability distribution over those eigenvectors.

Roger Penrose has suggested that this is sufficient reason to relax the ordinary dogma about self-adjoint (Hermitian) observables:
\begin{quote}
  ``In my opinion, this Hermitian requirement on an observable $Q$ is an unreasonably strong requirement, since complex numbers are frequently used in classical physics.... Since I am happy for the results of measurements (eigenvalues) to be complex numbers, while insisting on the standard requirement of orthogonality between the alternative states that can result from a measurement, I shall demand only that my quantum `observables' be normal linear operators, rather than the stronger conventional requirement that they be Hermitian.'' \citep[p.539]{penrose-roadtoreality}
\end{quote}
A similar suggestion has also been proposed by \citet{levyleblond1976a}, who pointed out that since a self-adjoint operator has spectral decomposition $A = \sum_{i}\lambda_i E_i$, every Borel function $f$ of a self-adjoint operator does too.

It turns out that a slightly stronger argument can be given, that there is in fact a sense in which the self-adjoint and normal operators are `projectively equivalent'. Let me turn to that statement now.

\subsection{Projective equivalence} 

The discussion above illustrates that both normal and self-adjoint operators allow us to apply the statistical interpretation afforded by the spectral theorem. Let me now add that there is a stronger sense of `equivalence' between these two classes of observables. This comes from a corollary of the spectral theorem, that in fact these two sets of operators give rise to the very same sets of projection valued measures. 

The general structure of an isolated quantum system can be characterised by a lattice of projections\footnote{This programme was initiated by \citet{mackey1963qm}; see \citet{redei-qm} for an introduction. I say `isolated' because physical features of subtly interacting systems like global phase are neglected the lattice description \citep{lyre2014a}.}. This lattice encodes the structural facts about a quantum system, including its commutation and uncertainty relations, as well as its possible quantum states. So, if the observables of two quantum systems share the same lattice of projections, then there is a strong sense in which they are equivalent descriptions. It turns out that normal operators and self-adjoint operators are equivalent in exactly this sense: they are extensions of the same underlying lattice of projections. For the sake of generality, I will frame this statement in terms of von Neumann algebras\footnote{A \emph{von Neumann algebra} $\mathcal{M}$ on a Hilbert space $\HH$ is one containing the identity $I$ and such that if a sequence $A_1,A_2,A_3,\dots$ in $\mathcal{M}$ weakly converges, in that there is linear $A$ satisfying $\Inn{\psi,A_i\phi}\rightarrow\Inn{\psi,A\phi}$ for all $\psi,\phi\in\HH$, then $A\in \mathcal{M}$.}. But if the reader is unfamiliar with this language, then one can just think of a particularly familiar von Neumann algebra, which is the set of all bounded linear operators on a Hilbert space.

\begin{figure}[tbh]\begin{center}
    \includegraphics[width=0.6\textwidth]{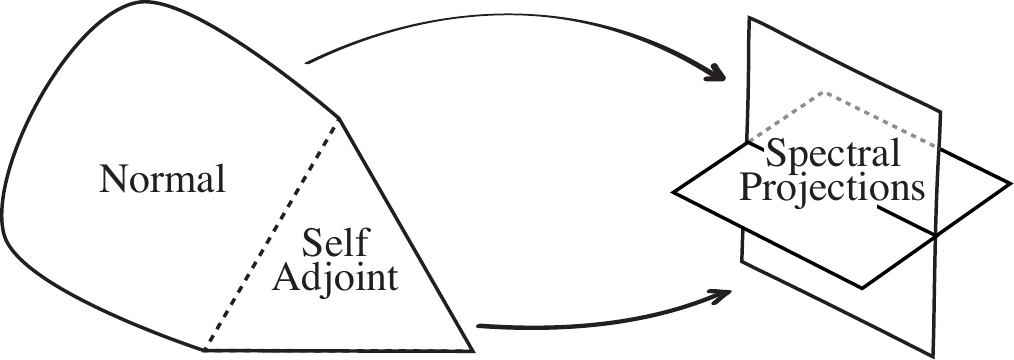}
    \caption{By Proposition \ref{prop:1}, the normal operators and the self-adjoint operators give rise to the same projective structure (and thus the same possible eigenvector decompositions) via the spectral theorem.}\label{fig:sa-projective}
  \end{center}\end{figure}

\begin{prop}\label{prop:1}
Let $\mathcal{M}$ be a von Neumann algebra on a Hilbert space with a countable basis. Let $\mathcal{N}\subseteq\mathcal{M}$ be the set of all its normal operators and $\mathcal{S}\subseteq\mathcal{M}$ the set of all its self-adjoint operators. Then, taking the spectral decompositions of $\mathcal{N}$ and of $\mathcal{S}$ produces the same lattice of projections.
\end{prop}
\begin{proof}
  Let $\mathcal{P}_{\mathcal{S}}$ denote the lattice of projections associated with the spectral decompositions for $\mathcal{S}$, and similarly $\mathcal{P}_{\mathcal{N}}$ for $\mathcal{N}$. Clearly $\mathcal{P}_{\mathcal{S}} \subseteq \mathcal{P}_{\mathcal{N}}$, since $\mathcal{S}\subseteq\mathcal{N}$. It thus remains to show that $\mathcal{P}_{\mathcal{N}}\subseteq\mathcal{P}_{\mathcal{S}}$. Like every linear operator, an element $A\in\mathcal{N}$ can be written $A = A_1 + iA_2$, with $A_1=(A+A^*)/2$ and $A_2 = (A - A^*)/2i$ both self-adjoint. Since $A$ is normal, it follows further that $A_1A_2 = A_2A_1$. So, $A$ is in the abelian von Neumann algebra generated by $A_1$ and $A_2$. But when acting on a Hilbert space with a countable basis, such an algebra always has a single self-adjoint generator $C$ \citep[Lemma II.2.8]{davidson1996a}. So, $A = f(C)$ for some Borel function $f$. Writing the spectral decomposition of this operator as $C = \int_{\mathbb{R}}\lambda dE_\lambda$, we thus find that $A = \int_{\mathbb{R}}f(\lambda)dE_\lambda$. Therefore, the unique projection valued measure $\Delta\mapsto E_\Delta$ associated with the spectral decomposition of $A$ is also associated with that of a self-adjoint operator $C$. Since $A$ was an arbitrary normal operator, it follows that $\mathcal{P}_{\mathcal{N}} \subseteq \mathcal{P}_{\mathcal{S}}$.
\end{proof}
Thus, from the `projective' perspective, normal operators provide a class of observables that just as reasonable as the class of self-adjoint operators. The situation is illustrated in Figure \ref{fig:sa-projective}.

\subsection{Symmetries and Dynamics}

A final question about normal operators as observables is how one ought to understand symmetries in this context. In orthodox quantum theory, there is a tight connection between symmetries and self-adjoint operators, which is reminiscent of Noether's theorem for variational symmetries. Namely, Stone's theorem guarantees a continuous group of symmetries is always generated by a unique self-adjoint operator. More precisely, if $s\mapsto U_s$ is a strongly continuous, one-parameter set of unitary operators satisfying $U_rU_s = U_{r+s}$ for all $r,s\in\mathbb{R}$, then there exists a unique self-adjoint operator $A$ such that $U_s = e^{isA}$ for all $s\in\mathbb{R}$  \citep[Thm. 5.9.2]{BlankExnerHavlicek}. Conversely, every self-adjoint operator generates a strongly continuous one parameter unitary representation of this kind. Examples: the spatial translation group  $a\mapsto U_a$ is generated by the momentum operator $P$, in that $U_a = e^{iaP}$. Similarly, the spatial rotation group $\theta\mapsto R_\theta$ is generated by the angular momentum operator $J$, in that $R_\theta = e^{i\theta J}$.

Can Stone's theorem be converted into an argument that observables must be self-adjoint operators? One might try to argue that continuous symmetries are generally associated with a conserved quantity, which we should think of as an observable. This does allow one to identify certain self-adjoint operators as observables. For example, the expectation value of momentum $P$ does not change under spatial translations, in that for any (pure or mixed) state represented by a density operator $\rho$, we have $\Tr{U_aPU_a^*\rho} = \Tr{U_aU_a^*P\rho} = \Tr{P\rho}$. However, this thinking works for normal operators, too: a whole host of normal operators are conserved along continuous unitary symmetries. Indeed, if $U_s$ is generated by the self-adjoint operator $A$, then every Borel function of $A$ is also similarly conserved, since such a function $f(A)$ always commutes with $U_s=e^{isA}$. As a result, non-self-adjoint normal operators like $iP$ and $e^{iaP}$ are conserved along spatial translations as well. So, conservation alone is no argument that observables are always self-adjoint operators. And after all, strictly speaking, the generator of a unitary group $U_s = e^{isA}$ is not even really a self-adjoint operator, but rather the `pure imaginary' operator $iA$, which has only imaginary numbers in its spectrum.

Still: even if there are non-self-adjoint normal observables, one might still insist on an ordinary unitary dynamics, which requires a self-adjoint generator $H$ (the `Hamiltonian'). The reasoning can be made precise as follows. Reflecting on our experience of time's passage, we might presume that time evolution is strongly continuous, as best we can tell. Isolated systems also seem to allow the same experiment to repeated at later moments in time with the same probabilistic outcomes, which is to say that the dynamics $t\mapsto U_t$ seems to satisfy time-translation invariance, $U_{t_1 + t_2} = U_{t_1}U_{t_2}$, with $U_t$ unitary so as to preserve probabilities\footnote{Unitarity in this sense for a continuous group follows by Wigner's theorem. One can alternatively presume that symmetries should preserve the norm of each state, since quantum states can be represented by normalised vectors; a corollary of Wigner's theorem then implies unitarity as well. I thank Adam Caulton for pointing this latter strategy out.}. Finally, suppose we presume that dynamical evolution holds (or could in principle hold) infinitely to the future and to the past, i.e. it can be described for all $t_1,t_2\in\mathbb{R}$. If one believes these things about the evolution of a quantum system, then Stone's theorem guarantees that the Hamiltonian observable $H$ is self-adjoint.

This perspective is certainly compatible with non-self-adjoint observables that are not the Hamiltonian. However, there may also be physical circumstances in which one or more of these presumptions assumptions fails. This can lead to the failure of unitarity and a failure of the Hamiltonian to be self-adjoint. For example, a non-isolated system does not satisfy the requirement of time translation invariance; we will discuss this example further in the discussion of radioactive decay in Section \ref{sec:wilderness}.

It may also be unreasonable to assume that dynamical evolution holds forever to the future and to the past. Such an assumption is much stronger than what is normally required of classical Hamiltonian mechanics, where only local time evolution is guaranteed\footnote{More formally: A smooth function $h:M\rightarrow\mathbb{R}$ on a symplectic manifold generates a Hamiltonian vector field, for which one can find a unique set of integral curves in a neighbourhood of each point. But it is perfectly possible for this Hamiltonian vector field that is incomplete, which is to say that its set of integral curves $\varphi(t)$ cannot be defined for all parameter times $t\in\mathbb{R}$.}. One might similarly expect that for some quantum systems, time translation might only be defined locally, perhaps because the system has a finite past, a finite future, or for some other reason altogether. This dynamical evolution will be generated by a Hamiltonian that is not-self-adjoint. Indeed, we will see explicit examples of this kind of evolution among the non-normal operators of the next section.

\section{Non-normal operators as observables}

Unlike normal operators, treating non-normal operators as observables extends quantum theory. It is not a mere adjustment of convention. Following the mathematical discussion above, there are three kinds of non-self-adjoint operators in this class: those that are symmetric but do not have a real spectrum, and those that have a real spectrum but are not symmetric, and those that satisfy neither. One may therefore choose exactly one of these commitments, or else reject them both. The first commitment allows one to extend quantum theory to include time observables; the second kind includes so-called PT-symmetric quantum mechanics; the third allows for these and yet other possible physical descriptions. I will provide a little overview of each below, arguing that there are circumstances in which each is reasonable.

\subsection{Symmetric operators and time observables}

Let me begin this section by recalling a case in which one treats a non-self-adjoint operator like an observable, but only because it can be extended to a self-adjoint operator. I will then turn to the more important case for my purposes, of non-self-adjoint operators that cannot be extended in this way.

\begin{figure}[tbh]\begin{center}
    \includegraphics[width=0.5\textwidth]{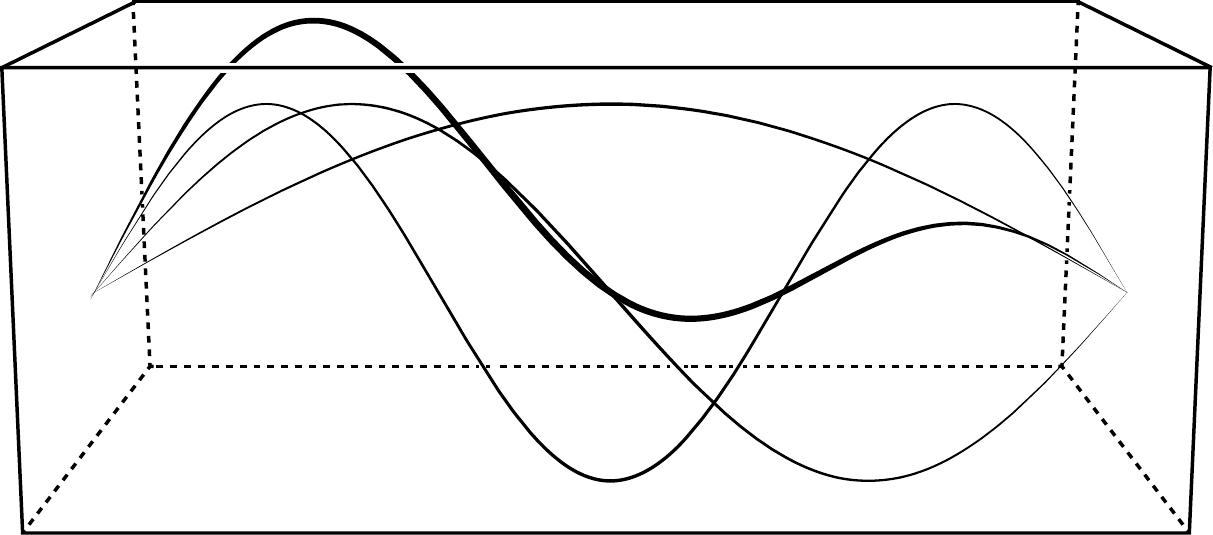}
    \caption{$P=i\tfrac{d}{dx}$ is not self-adjoint on differentiable wavefunctions that vanish at the sides of a box, $\psi(a)=\psi(b)=0$.}\label{fig:wavefunctions}
  \end{center}\end{figure}

\emph{Case 1: Self-adjoint extensions}. Suppose we wish to describe a particle in a box of finite width $b-a$. We adopt the Hilbert space of square-integrable wavefunctions $\psi(x)\in L^2([a,b])$ with $\Inn{\psi,\phi} := \int_a^b\psi^*(x)\phi(x)dx$. As experienced quantum mechanics, we wish to see a momentum observable for this particle that looks like the standard momentum operator $P=id/dx$. Such an unbounded operator cannot act on the entire Hilbert space. The art of unbounded operators is thus to answer the question: which wavefunctions does the operator it act on? Since it also has different properties depending on the domain acted upon, let me for the moment describe momentum as an operator-domain pair $(P,D)$. Suppose we identify the domain $D$ as the set of differentiable functions that vanish at the edges of the box, $\psi(a)=\psi(b)=0$, as shown in Figure \ref{fig:wavefunctions}. Call this domain $D_0$. Then $(P,D_0)$ can be shown to be closed, densely-defined and symmetric; however, it is not self-adjoint \citep[Example 4.2.5]{BlankExnerHavlicek}. It is also non-normal and fails to have a purely real spectrum, as a consequence of our mathematical discussion above.

Nevertheless, it is common practice to view non-self-adjoint operators like this one as observables. This is because one can turn $(P,D_0)$ into a self-adjoint operator by extending its domain again, this time to include all the wavefunctions that satisfy $\psi(a)=e^{i\theta}\psi(b)$ for some fixed real $\theta$. Call this extended domain $D_\theta\supset D_0$. Then $(P,D_\theta)$ is self-adjoint, for each real number $\theta$ (\emph{ibid}). It is thus common practice to construct a symmetric observable that is not self-adjoint, with the aim of extending them to a self-adjoint operator as needed. As \citet{earman2009essential} has pointed out, this is a particularly hazardous practice when the observable being considered is the Hamiltonian. For, as in the case of the particle in a box, there may be multiple self-adjoint extensions of a symmetric observable. Symmetric operators with a unique self-adjoint extension are called \emph{essentially self-adjoint}. When a Hamiltonian fails to have this property then it may have multiple distinct self-adjoint extensions. This gives rise to multiple unitary evolutions, and therefore a failure of determinism for Schr\"odinger evolution. The failure of determinism might lead one to be skeptical of treating non-self-adjoint operators with multiple self-adjoint extensions as observables. This is not so convincing if one takes the question of determinism to be an open one\footnote{\citet[p.36]{earman2009essential} still supports the practice of treating some symmetric operators that are not self-adjoint as observables, calling its rejection ``high handedness''; \citet[p.373]{wuethrich2011b} agrees for this reason that ``the question of whether the Schr\"odinger evolution is deterministic does not afford a simple and unqualified answer."}. But another class of non-self-adjoint symmetric operator is even more convincing. That class is the following.

\emph{Case 2: Maximal symmetric operators} Let me now turn to  cases that may appear even worse, but are in fact better: symmetric operators that do not admit any self-adjoint extensions at all. Such operators are called \emph{maximal symmetric}. If we wish to treat a maximal symmetric operator like an observable, then it cannot stand proxy for a self-adjoint extension; it must be treated like an observable in its own right. 

The assumptions of Stone's theorem fail for maximal symmetric operators, so they do not generate a unitary group in the usual sense. However, they do satisfy a closely related result. Stating this result makes use of the concept of an \emph{isometry}, a Hilbert space operator $U$ satisfying $U^*U\psi=\psi$ for all $\psi$ in its domain (a \emph{unitary} operator is an isometry that is bijective). Isometries are symmetry transformations in much the same sense as a unitary operator, except that they are not defined on all states. They also allow us to state the following generalisation of Stone's theorem\footnote{This result follows naturally from the work of \citet{neumark1940a,naimark1968linear} on the theory of self-adjoint extensions, although it was proved independently by \citet{cooper1947,cooper1948symmetric}.}. 

\begin{genstonethm}
If $s\mapsto U_s$ is a strongly continuous, one-parameter set of isometries satisfying $U_rU_s = U_{r+s}$ for all $r,s\geq 0$ (or for all $r,s\leq0$, but not both), then there exists a unique maximal symmetric operator $A$ such that $U_s = e^{isA}$. Conversely, every maximal symmetric operator $A$ generates a strongly continuous one parameter set of isometries set $s\mapsto U_s=e^{isA}$ satisfying $U_rU_s = U_{r+s}$, for all $r,s\geq 0$ (or for all $r,s\leq 0$, but not both). \citep{cooper1947,cooper1948symmetric}
\end{genstonethm}
This means that maximal symmetric operators are associated with a set of symmetries after all, in very much the same way as self-adjoint operators. These symmetries are simply limited to a restricted domain, in addition to be limited by the parameter values of the set.

When a maximal symmetric observable is a Hamiltonian, the Generalised Stone Theorem says that a unique solution to the Schr\"odinger equation exists, although it is only defined for non-negative times or non-positive times (but not both). As far as determinism is concerned, this situation is an improvement on the failure of essential self-adjointness considered by \citet{earman2009essential}. The generalised Stone theorem says that the dynamical evolution generated by a maximal symmetric Hamiltonian is unique, much like the dynamics of an essentially self-adjoint Hamiltonian. The dynamics is time-translation invariant, in the restricted sense of an isometry. The limitation is just that this dynamics is not defined for all times $t\in\mathbb{R}$. As discussed above, having a dynamics for all times is a very strong requirement, which we may have good reason to relax.

Maximal symmetric operators also fail to satisfy the conditions of the ordinary spectral theorem. But there is an interesting generalisation of this too, which makes use of Positive Operator Valued Measures (POVMs) on Borel sets of the reals. A POVM generalises our earlier notion of a projection valued measure, by carrying over its properties exactly but for positive operators instead of projections\footnote{A positive operator is one that has a non-negative spectrum, $\mathrm{Sp}(A)\geq0$.}. Such measures allow us to state the following.

\begin{genspecthm}
Let $A$ be a closed, densely defined symmetric operator. Then there exists a POVM $\Delta\mapsto F_\Delta$ such that $A = \int_{\mathbb{R}}\lambda dF_\lambda$, which is unique (up to unitary equivalence) if and only if $A$ is maximal symmetric, and which is a Projection Valued Measure if and only if $A$ is self-adjoint. \citep[Thm. 5.16, pg.135]{DubinHennings1990a}
\end{genspecthm}
Just like self-adjoint operators, maximal symmetric operators have a unique spectral decomposition. It is just not in terms of a projection valued measure. As \citet*{BuschEtAl1994,BuschGrabow1995a} have pointed out, a POVM and a state still give rise to a probability distribution, which allows one to interpret a POVM statistically in an experiment. However, the details of that interpretation can be subtle. A curiosity about POVMs is that two elements $F_\Delta$ and $F_{\Delta^\prime}$ with $\Delta\cap\Delta^\prime=\varnothing$ are not necessarily orthogonal, or even commutative. Consequently, given a state $\psi$ and a maximal symmetric operator $A$ with a pure point spectrum, the decomposition $\psi = c_1\varphi_1 + c_2\varphi_2 + \cdots$ may be in terms of eigenvectors $\varphi_i$ of $A$ that span the Hilbert space, but which are not all pairwise orthogonal. Such a POVM is sometimes interpreted as arising out of some uncontrolled aspect of a measurement procedure \cite[\S 2.2.6]{NielsenChuang2000qi}. But it can also occur in more elementary measurement scenarios.

I will just mention one such scenario to illustrate, which is the case of `time observables'. A time observable is a natural object of study when one is interested in durations of some event, or in the time that something occurs. For example, one might wish to use an observable to describe the moment that a particle in flight arrives at its target, or the time that a jet or particle decay occurs in a detector. Such time observables are a natural candidate for description in terms of maximal symmetric operators. For such observables, there is no need physical need to require the eigenstates to be pairwise orthogonal; after all, the time that something occurs is not statistically independent of previous times.

There is a literature that has come to this same conclusion through another route. Let $\HH$ be a Hilbert space, together with an ordinary unitary dynamics defined by $t\mapsto U_t$. We call a linear operator $T$ a \emph{time operator} if and only if it satisfies $U_t TU_t^* = T + tI$ for all $t\in\RR$. Equivalently, for any $\psi\in D_T$ with $|\psi|=1$ and $\psi(t) := U_t\psi$, a time operator $T$ is one that satisfies $\Inn{\psi(t),T\psi(t)} = \Inn{\psi,T\psi} + t$, for all $t\in\RR$. These properties can be informally summarised as requiring that a time observable `tracks' the evolution of time determined by the unitary dynamics. An operator $T$ with this property is in general unbounded, and also satisfies the time-energy commutation relation $[H,T]=i$, which is its `local' expression.

The central no-go result for time operators, known as \emph{Pauli's theorem}, is that if $U_t=e^{-itH}$ is generated by a Hamiltonian $H$ that is bounded from below (as almost all known Hamiltonians are), then every time operator fails to be self-adjoint\footnote{This result is inspired by a famous remark of \citet[pg.63, fn.2]{pauli-qm}, which was made more rigorous e.g. by \cite[\S VII.6]{ludwig1983a} and \citet{srinivas1981time}, among others.}. This fact was originally interpreted as an impossibility result for time observables, and is sometimes referred to as the `problem of time' in quantum mechanics \citep{butterfield2013time}. However, if we relax our requirements on what counts as an observable, then it can equally be viewed as simply saying that time operators are non-self-adjoint observables. Then there turn out to be a plethora of possible time observables, most known examples of which are maximal symmetric. 

A particularly simple time observable\footnote{This example was identified by \citet{aharonov1961time}. For further discussion, see also \citet{holevo1982a,BuschEtAl1994,galapon2009pauli,pashby-2014thesis}.} can be seen for the free particle Hamiltonian $H = \tfrac{1}{2m}P^2$. This dynamical system admits a time operator given by,
\[
  T = \tfrac{m}{2}(QP^{-1} + P^{-1}Q).
\]
This $T$ is a time operator because the free particle satisfies $e^{-itH}Qe^{itH}=Q + \tfrac{t}{m}P$ and $e^{-itH}Pe^{itH}=P$, from which it easily follows that $e^{-itH}Te^{itH}=T + tI$. It is also symmetric by construction.

However, we can immediately infer from Pauli's theorem that this time operator is not self-adjoint, and with a little more work show that it does not have any self-adjoint extensions \citep[\S 3.8]{holevo1982a}. It follows that $T$ is maximal symmetric. A large class of dynamical systems with such time operators has been constructed by \citet{BuschEtAl1994,HegerfeldtMuga2010a}, and these observables have been put to many interesting uses \citep{MugaEtAl2008}. A closely-related discussion exists for `phase observables' as well \citep{BuschGrabow1995a}.

Let me summarise the discussion of this section. The addition of maximal symmetric operators as observables is a non-trivial extension of quantum theory. However, it is a mathematically controlled extension, thanks to a generalised Stone theorem and spectral theorem. These generalisations introduce features that are unfamiliar from the perspective of more traditional observables. However, even these unfamiliar features can be made sense of in concrete physical descriptions in which we can put maximal symmetric operators to use. Little reason remains to deny their status as bona fide `observables'.  

\subsection{Real spectrum operators and PT symmetry}


Among the most commonly demanded requirements on a quantum observable is that it should have a real spectrum. Non-self-adjoint operators with a real spectrum thus provide another natural route to extending observables in quantum mechanics. However, as we shall see, this class of operators is much more unwieldy that the previous ones, with no analogue of the spectral theorem nor of Stone's theorem without adding extra structure to the theory.

We have discussed the matrix $A = \smmatrix{1 & 1}{0 & 2}$ as an example of a non-self-adjoint operator with a real spectrum. A much more interesting example from the perspective of physical applications is the operator,
\begin{equation}\label{eq:bender}
  H = \tfrac{1}{2m}P^2 + \tfrac{m\omega^2}{2}Q^2 + iQ^3,
\end{equation}
where $Q$ and $P$ are the position and momentum operators in some representation of the canonical commutation relations, and $m,\omega\in\mathbb{R}^+$. This operator is obviously not symmetric, and therefore fails to be self-adjoint or even normal. However, has been studied extensively following the work of \citet{BenderBoettcher1998a} as a possible interaction Hamiltonian, and was proven  by \citet{dorey2001spectral,dorey2001supersymmetry} to have an entirely real spectrum, with interesting connections to supersymmetry. More general classes of non-self-adjoint operators with this property have also been explored, by considering a class of $\mathcal{PT}$-symmetric operators \citep{BenderEtAl2002a,BenderEtAl2003a}, or more generally those that commute with an antilinear operator \citep{weigert2003completeness}.

It is possible to identify a general class of operators that have a real spectrum, but are not necessarily self-adjoint. Following \citet{streater-lostcauses}, call a linear operator $A$ a \emph{diagon} if and only if there exists an operator $B$ with densely-defined inverse such that $BAB^{-1}$ is self-adjoint. Since the similarity transformation $(\cdot)\mapsto B^{-1}(\cdot)B$ is spectrum-preserving, it follows that every diagon has a real spectrum. In particular, a self-adjoint operator is a diagon with $B=I$ the identity. However, diagons are certainly not always self-adjoint. For example, if $Q$ and $P$ are the position and momentum operators in the Schr\"odinger representation, then $Q^{-1}PQ$ is a non-self-adjoint diagon. It can be transformed to the self-adjoint operator $P$ by the similarity transformation $(\cdot) \mapsto Q(\cdot)Q^{-1}$, and thus has a real spectrum. But it is easy to check that it is not symmetric, not normal, and therefore not self-adjoint\footnote{Apply the commutation relations to see that it is not symmetric: $(Q^{-1}PQ)^* = QPQ^{-1} = iQ^{-1} + P$, whereas $Q^{-1}PQ = iQ^{-1} - P$. One can use these facts to check that $Q^{-1}PQ$ is also not normal.}.

One strange feature of non-self-adjoint diagons is that their expectation values may not be real, even though the spectrum of the operator is. In the example above, this is easy to check: for an arbitrary vector $\psi$ in the common domain of $Q$, $P$ and $Q^{-1}$, we have by application of the commutation relations that,
\[
  \Inn{\psi,(QPQ^{-1})\psi} = \Inn{\psi,(iQ^{-1} + P)\psi} = i\Inn{\psi,Q^{-1}\psi} + \Inn{\psi,P\psi}.
\]
This implies that $QPQ^{-1}$ has expectation values with a pure-imaginary component. \citet[\S 12.5]{streater-lostcauses} has pointed out that such complex expectation values are quite general features of non-self-adjoint diagons. Thus, even though these operators retain what many have taken to be the `gold standard' of observables, a real spectrum, the are quite difficult to interpret by themselves in the context of ordinary quantum theory.

A non-self-adjoint operator with a real spectrum is never `diagonalizable' in the usual  sense: it does not have a spectral decomposition in the sense of a projection valued measure, since the ordinary spectral theorem applies only to normal operators. It is not known whether a more general spectral theorem exists for such operators, analogous to the Naimark spectral theorem for symmetric operators. The application of Stone's theorem suffers from similar difficulties.

However, the spectral structure of a large class of real-spectrum operators has been studied using other kinds of decompositions, introduced by \citet{bender2000conjecture} and developed by \citet{weigert2003completeness} and others. The usual statistical interpretation of quantum theory is not possible for these operators, since they do not admit a projection-valued measure. However, an interpretation is still possible if one introduces a new inner product, and then defines the statistics and the dynamics with respect to that. For example, if $H$ is a non-self-adjoint diagon on a Hilbert space with inner product $\Inn{\cdot,\cdot}$, then one can always construct a new inner product $\Inn{\cdot,\cdot}_H$ with respect to which $H$ is self-adjoint\footnote{This is a straightforward exercise: first show that if $H$ is a diagon with respect to $\Inn{\cdot,\cdot}$, then it is `quasi-Hermitian' with respect to $\Inn{\cdot,\cdot}_B := \Inn{\cdot,B(\cdot)}$, meaning that $BHB^{-1}=H^*$. Then show that if $H$ is quasi-Hermitian with respect to $\Inn{\cdot,\cdot}$, then it is self-adjoint with respect to $\Inn{\cdot,\cdot}_B$.}. One can then take spectral decompositions and define a unitary dynamics in the resulting new Hilbert space. This strategy, proposed by \citet{BenderEtAl2002a}, has been the subject of a great deal of fruitful research\footnote{For overviews, see \citet{bender2007making,moiseyev2011non,znojil2015a}.}.

We thus have an interesting extension of quantum theory, in which we only require the spectrum of each observable to be real, and each non-self-adjoint observable requires its own inner product. However, a central requirement of this programme, that one must always have real eigenvalues, is perhaps not as clear as it should be. We need not place so much weight on states represented by real numbers. The previous sections have reviewed many scenarios in which non-real numbers can be used to represent physical experiments. If one is willing to relax the requirement of self-adjointness at all, then one should minimally allow for complex eigenvalues, too. This is the topic of the last section.

\subsection{None of the above: The wilderness beyond}\label{sec:wilderness}

For each of the properties of being normal, being symmetric, and having a real spectrum, there is a literature on retaining that property while giving up other two. In this section, we discuss the possibility of giving up all three. Then there is no single mathematical idea controlling the concept of an observable, which leads to a loss of generality of the usual mathematical results that more traditional observables enjoy. However, the result is not necessarily a complete free-for-all. There are physical ideas that allow even these operators to be interpreted as observables, and mathematical results that allow us to control their behaviour. 

\begin{figure}[tbh]\begin{center}
    \includegraphics[width=0.5\textwidth]{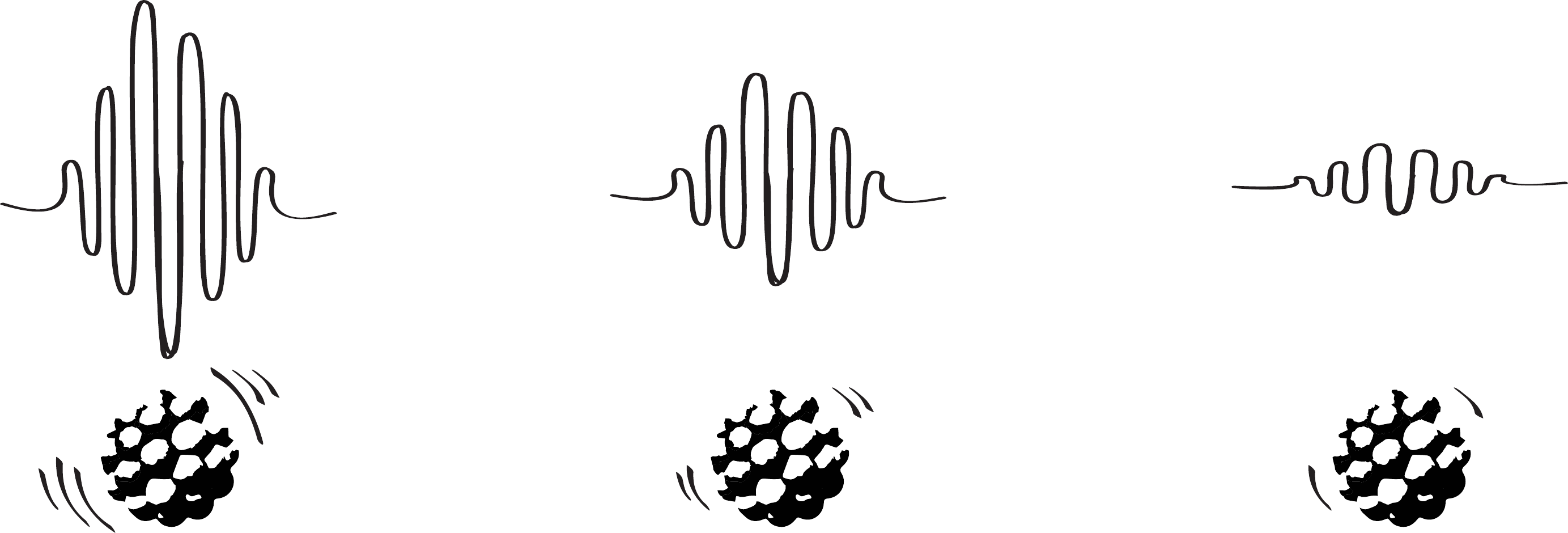}
    \caption{Gamow's (1928) model of radioactive decay used a non-self-adjoint Hamiltonian with eigenvalue $E = E_0 - i\Gamma$.}\label{fig:nonunitary-decay}
  \end{center}\end{figure}

An early example of such an observable was proposed by \citet{gamow1928tunneling}, in a famous paper written on a visit to G\"ottingen that introduced the world to quantum tunneling. Adopting Schr\"odinger's wavefunction formalism, Gamow proposed that the energy value of a radioactive particle could be described by a complex number,
\[
  E = E_0 - i\Gamma.
\]
He gave an immediate physical interpretation of this value, identifying  $E_0$ the `ordinary energy' and $\Gamma$ as a positive `damping term'. In particular, a corresponding energy eigenstate $\phi$ would evolve according to the rule,
\[
  \phi(t) = e^{-itE}\phi = e^{-it(E_0 -i\Gamma)}\phi =e^{-t\Gamma}e^{-itE_0}\phi.
\]
This state is nearly stationary $\phi(t)\approx e^{-ita}\phi$ when $t\approx0$, but has the decreasing amplitude of a damped wave with decay rate $\Gamma$, as shown in Figure \ref{fig:nonunitary-decay}.

What sort of observable generates the dynamics $t\mapsto e^{-itH}$ for this system? It is a non-self-adjoint Hamiltonian,
\[
  H = A - iB,
\] 
where $A$ and $B$ are self-adjoint operators each with a positive spectrum. The dynamics fails to be unitary because it is a system that is a non-isolated system, which subtly interacts with its environment. The operators $A$ and $B$ typically do not commute, and so the Hamiltonian $H$ is not generally normal. It is also non-symmetric, and has a non-real eigenvalue $E=E_0-i\Gamma$ by construction. So, the Hamiltonian for Gamow's quantum tunneling system is one that fails all the criteria for observables that we have discussed so far. This example was discussed in an influential textbook by \citet[pgs.555-556]{LandauLifshitz1958qm}, and has given risen to literatures that use non-self-adjoint Hamiltonians to describe quantum resonance and quantum optics \citep[see e.g.][]{moiseyev2011non}.


Although there are many other operators that are non-symmetric, non-normal, and have a non-real spectrum, it is not always easy to assign them a physical interpretation. For example, if $\sigma_x$, $\sigma_y$, $\sigma_z$ are the standard Pauli spin matrices, then we can formally write down the operator,
\[
  R=\sigma_x + i(\sigma_y + \sigma_z).
\]
This operator has complex eigenvalues $\pm i$, associated with eigenvectors $\smmatrix{i}{1}$ and $\smmatrix{0}{1}$, respectively. It has an overcomplete basis just like the maximal symmetric operators, but is non-symmetric and non-normal. Unfortunately, the physical significance of such an operator is also far from clear. It certainly does not admit an obvious interpretation as the generator of a dynamics.

These are just a few examples from the wilderness of non-self-adjoint operators. Much remains to be learned about the structural properties of such operators, such as their spectral theory and physical applications, and research in this area is ongoing. But this should not prevent us from exploring their possible use as observables.

\section{Conclusion}

In this paper we have sorted non-self-adjoint operators into four classes: those that are normal, those that are symmetric, those that have a real spectrum, and those that admit none of these properties. In spite of a pervasive dogma, non-self-adjoint operators may provide conceptual clarity or calculational convenience in modeling quantum systems. We have seen that the first class, that of normal operators, is in many senses equivalent to standard quantum mechanics. In contrast, the second and third classes introduce varying degrees of new physics into the discussion, from time observables to new interaction Hamiltonians. The fourth class is a wilderness of many unknowns. But some of them can be used to fruitfully model quantum systems.

In his textbook on linear operators, E. Brian Davies gave an apt characterisation of the state of non-self-adjoint operators from a mathematical perspective:
\begin{quote}
  Studying non-self-adjoint operators is like being a vet rather than a doctor: one has to acquire a much wider range of knowledge, and to accept that one cannot expect to have as high a rate of success when confronted with particular cases. \citep[p.x]{davies2007book}
\end{quote}
So too is the proposal to allow observables that are not self-adjoint. When an arbitrary non-self-adjoint operator is proposed as an observable from the great wilderness of possibilities, there may well be little that we can say about how to associate it with real-world observations. However, a number of interesting cases are well-understood, philosophically well-motivated, and lead to physically relevant models of quantum theory. It would be a pity if mere dogma prevented us from enjoying them.

\end{document}